\documentstyle[11pt] {article}
\newlength{\mytopmargin}
\newlength{\myleftmargin}
\begin{document}
\title{
Exact Calculation of the Ground State \\
Single-Particle Green's Function for the $1/r^2$ \\
Quantum Many Body System at Integer Coupling}
\author{P.J. Forrester\thanks{e-mail: matpjf@maths.mu.oz.au}
\\
Department of Mathematics \\
University of Melbourne \\
Parkville, Victoria  3052 \\
Australia}
\date{}
\maketitle
\begin{abstract}
The ground state single particle Green's
function
describing hole propagation is calculated exactly
for the $1/r^{2}$ quantum many body system at integer
coupling.  The result is in agreement with a
recent conjecture of Haldane.
\end{abstract}

\section{Introduction}
\setcounter{equation}{0}
\renewcommand{\theequation}{1.\arabic{equation}}

The $1/r^{2}$ quantum many body problem, and the closely related
Haldane-Shastry $1/r^{2}$ Heisenberg chain, are the subject of intense
theoretical study at present, due in part to their relationship to an ideal
gas obeying fractional statistics (see [1] for a review).  One direction of
study has been the exact calculation of some ground state correlation
functions [1-5]. Let us briefly summarize the main results of these works.

The static ground state correlations (one-body density matrix, two-point
distribution function) for the $1/r^{2}$ Hamiltonian in periodic boundary
conditions
\begin{equation}
H:=-\sum_{j=1}^N \; \frac{\partial^{2}}{\partial x^{2}_{j}} +
2q(q-1) \left(\frac{\pi}{L}\right)^{2} \sum_{1 \leq j<k \leq N}
\frac{1}{\sin^{2}
\pi (x_k-x_j) /L} ,\vspace{.2cm} q \in Z^+
\end{equation}

\noindent (here we have set $\hbar^2/2m :=1$), for which the exact ground
state wave function is
\begin{equation}
| \; 0 > := \psi_0 (x_1, \ldots ,x_N) := \prod\limits_{1 \leq j<k \leq N}
\Big( \sin \pi (x_k-x_j)/L \Big) ^q \; ,
\end{equation}

\noindent have been calculated exactly in terms of a class of generalized
hypergeometric functions in several variables based on the Jack symmetric
polynomials [2].  These functions are defined in terms of power series, and
in the cases of interest an explicit expression for each coefficient is
known.  However, the coefficient of the $N$th term involves a sum over all
partitions of $N$ into $q$ parts and is thus impractical to compute except for
small $N$ and $q$.  To overcome this difficulty, recurrence formulas have been
presented [3] which allows for the rapid numerical evaluation of the ground
state correlations in the finite system.  Furthermore, integral
representations of the generalized hypergeometric functions
are known [4,5], from which the density matrix and two-point distribution
function have been written in terms of $q$ and $2q$ dimensional integrals
respectively.

The thermodynamic limit can be taken in the integral formulas, giving
similar integral formulas for the density matrix and two-point distribution
function in the thermodynamic limit [5].  This integral formula for the
density matrix has recently been generalized by Haldane [1] to an integral
formula for the retarded single-particle Green's function
\begin{equation}
-iG \Big( (x,t),(0,0) \Big) :=TL<0 \mid \psi^{+} (x,t) \psi (0,0) \mid 0>
\end{equation}

\noindent where TL denotes the thermodynamic limit $N,L \rightarrow \infty,
N/L= \rho$.  Haldane {\it conjectures} that for general $q \in Z^+$

$$
-iG \Big( (x,t),(0,0) \Big) = \frac{\rho}{2} \: B_{q}
\int\limits_{[-1,1]^q} dv_1 \ldots dv_q \prod^q_{j=1} (1-v^2_j)^{-1+1/q}
\hfill
$$
\begin{equation}
\times \prod_{1 \leq j<k \leq q} |v_k - v_j|^{2/q} \prod^q_{j=1}
e^{i \pi \rho x v_j} e^{-iq(\pi \rho)^2 (1-v^2_j) t/ \hbar}
\hspace{1cm}
\end{equation}
\\ [.1in]

\noindent where
\begin{equation}
B_{q}:= \prod^q_{j=1} \; \frac{\Gamma (1+1/q)}{\Big(\Gamma (j/q) \Big)^2} =
\frac{q}{(2 \pi )^{q-1}} \Big(\Gamma (1+1/q) \Big)^q
\end{equation}
\\ [.1in]
\noindent As $t \rightarrow 0^+$ the results of [5] for the density matrix
are reclaimed.  For general $t$, this result was proved in [6] for $q=2$
using a mapping to a dynamical matrix model; for $q=1$ - free fermions - it
is easy to verify from the differential equation satisfied by
$-iG \Big( (x,t),(0,0) \Big)$ (see e.g. [7]).

\section{Derivation of Haldane's Conjecture}
\setcounter{equation}{0}
\renewcommand{\theequation}{2.\arabic{equation}}
\subsection{Action of the Hamiltonian on a hole state}

Our starting point is to write the retarded single-particle Green's function
for the finite system in a more explicit form (see [6], eq. (12):
$$
-iG_{N+1} \Big( (x,t);(0,0) \Big) = (N+1) < 0 \bigm| \prod^N_{l=1} (\sin \pi
(x_l-x)/L \Big)^q e^{-i(H-E_0)t/ \hbar} \hfill
$$
\begin{equation}
\hspace{2.3in} \times \prod\limits^N_{l=1} (\sin \pi x_l/L)^q \bigm| 0 >
\hspace{.3in}
\end{equation}

\noindent Next we introduce the auxiliary variables $y_1, \ldots ,y_q$ and
take up the task of calculating
\begin{equation}
(N+1)<0 \bigm| \prod^N_{l=1} \: \prod^q_{j=1} \sin \pi (x_l-y_j)/L \,
e^{-i (H-E_0)t/ \hbar} \: \prod^N_{l=1} (\sin \pi x_l/L)^q \bigm| 0 >.
\end{equation}

\noindent We do this by first expanding the exponential

\begin{equation}
e^{-i(H-E_0)t/ \hbar} = \sum^{\infty}_{j=0} \frac{(-it/ \hbar)^j}{j!}
\; (H-E_0)^j
\end{equation}

\noindent and considering the action of the operator $H-E_0$ on the hole state
\begin{equation}
|  \phi > := \prod^q_{j=1} \sin \pi (x_l-y_j)/L \bigm| 0 >.
\end{equation}

\noindent We have the following result.

\vspace{.2cm}
\noindent \underline{Lemma 1} \\
\noindent With $H$ given by $(1.1)$, $E_0$ the corresponding ground state
energy and $| \: 0>$ given by (2.4) we have
$$
(H-E_0) \bigm| \phi > \; = T_{\{ y\} } \bigm| \phi >
\eqno (2.5a)
$$
where
$$
T_{\{ y\} } := q \left[ \; \sum^q_{j=1} \:
\frac{\partial^2}{\partial y^2_j} + {\frac{\pi}{qL}}^2\sum^q_{j_1 \neq j_2}
\cot
\pi (y_{j_1}-y_{j_2})/L \right. \hfill
$$
$$
\times \left. \left( {\partial \over \partial y_{j_1}}-{\partial \over \partial
y_{j_2}} \right) + \left ( {\pi \over L} \right )
\Big( qN+N(N-1) \Big) \right]
\eqno (2.5b)
$$

\setcounter{equation}{5}

\noindent \underline{Proof} \\
\noindent Direct differentiation gives
$$
(H-E_0) \bigm| \phi > = \hspace{5cm}
$$
$$
 - \left(\frac{\pi}{L}\right)^2 \left[ -qN+
\sum^N_{l=1} \; \sum^q_{j_1 \neq j_2} \cot \pi (x_l-y_{j_1})/L \cot
\pi(x_l-y_{j_2})/L \right.
$$
\begin{equation}
+\left. 2q \sum^N_{l=1} \; \sum^N_{l'=1 \atop l' \neq l} \cot \pi
(x_l-x_{l'})/L \:
\sum^q_{j=1} \cot \pi (x_l-y_j)/L \right] \bigm| \phi >
\end{equation}

\noindent We rewrite the second summations by grouping together the summand
with the summand with $l$ and $l'$ interchanged.
Simple manipulation gives
$$
\cot \pi (x_l - x_{l'})/L \left[ \cot \pi (x_l - y_j )/L  -
\cot \pi (x_{l'} - y_j) /L \right] \hfill
$$
\begin{equation}
= -\left(  1 + \cot \pi (x_\l - y_j )/L \cot \pi (x_{l'} - y_j)
/L
\right ).
\end{equation}

\noindent In the first summations we use (2.7) to rewrite the summand:
$$
\cot \pi (x_l - y_{j_1})/L \cot \pi (x_l - y_{j_2})/L \hfill
$$
\begin{equation}
 = -1 - \cot \pi (y_{j_1} - y_{j_2})/L \big [
\cot \pi (y_{j_1} - x_l )/L - \cot \pi (y_{j_2} - x_l)/L \big ].
\end{equation}

\noindent Substituting (2.8) and (2.7) in the first and second summations
respectively of (2.6) we obtain
$$
(H - E_0) \bigm| \phi > \, = - \left ( \pi \over L \right )^2 \big [
-q^2 N- \sum_{j_1 \not= j_2}^q \cot \pi (y_{j_1} - y_{j_2}) / L  \hfill
$$
$$
 \times \sum_{l=1}^N \left ( \cot \pi (y_{j_1} - x_l)/L
- \cot \pi (y_{j_2} - x_l)/L \right ) - q N (N - 1)
$$
\begin{equation}
-q \sum_{j=1}^q \sum_{l_1 \ne l_2 }^N
\cot \pi (x_l - y_j) /L \cot \pi (x_{l'} - y_j)/L \big ]
\end{equation}
But from (2.4) and (1.2)
\begin{equation}
{\partial \over \partial y_j} \bigm | \phi >
= - \left ( \pi \over L \right ) \sum_{l = 1}^N \cot \pi (x_l - y_j)/L
\: \bigm |\phi >
\end{equation}
and
\begin{equation}
{\partial^2 \over \partial y_j^2} \bigm |\phi >
= \left (\pi \over L \right )^2 \sum_{l',l = 1 \atop l' \not= l}^N
\cot \pi (x_l - y_j)/L \: \cot \pi (x_{l'} - y_j)/L \bigm |\phi >.
\end{equation}
Substituting (2.10) in (2.9) we obtain (2.5), as required.
\hfill $\Box$

\vspace{.5cm}
\noindent Remark: The operator $T_{\{y\}}$ has occured previously in the study
of the $1/r^2$ quantum many body system [8]. Thus if $\psi$ denotes an
eigenfunction of $H$ with $q$ replaced by $1/q$ (recall (1)), and
$\psi = \Phi \bigm |0 >,$ then $\Phi$ is an eigenfunction of $T_{{y}}$.
Furthermore, it is known [4] that the eigenfunctions of $T_{{y}}$ are the
Jack polynomials $C_{\kappa}^{(q)} (e^{2 \pi i y_1 /L}, \dots ,e^{2 \pi i y_q
/L})$ where $\kappa$ is a partition which labels the eigenfunction. This
latter fact will play a crucial role in our subsequent analysis.
\vspace{.5cm}

Using Lemma 1 and (2.3), we see that
$$
 -i G_{N+1} \bigm ( (x,t),(0,0) \bigm ) \hfill
$$
\begin{equation}
= (N+1) \sum_{j=0}^\infty {(-it/\hbar)^j \over j!} \left (T_{\{y\}} \right )^j
< \phi \bigm | \prod_{l=1}^N \sin \pi x_l /L \bigm | \phi > \bigm |_{
y_1= \dots =y_q=x}.
\end{equation}
We have previously evaluated the inner product in (2.11), which at the
specified point is precisely the static one-body density matrix. Thus
[4, eq.(3.4)\footnote{this equation is erroneously missing the factor
$C_{q,N}$}]
$$
(N+1)<\phi \bigm | \prod_{l=1}^N (\sin \pi x_l /L)^q \bigm |0> = \hfill
$$
\begin{equation}
 \left ({N + 1 \over L} \right ) C_{q,N} \prod_{l=1}^q e^{-\pi i \rho y_l}
{_2F_1}^{(q)}(-N,1;-N+1-1/q;e^{2 \pi i y_1 /L}, \dots , e^{2 \pi i y_q /L} ),
\end{equation}
where
\begin{equation}
C_{q,N} = {A_q(1/q,-1,2/q) \over A_q(1/q,-N-1,2/q)}
\end{equation}
with
$$
A_n(\lambda_1,\lambda_2,\lambda) =
\prod_{j=1}^n {\Gamma (1 + \lambda /2) \Gamma (\lambda_1 + \lambda_2
+\lambda (n + j - 2)/2) \over
\Gamma (1 + \lambda j/2) \Gamma (\lambda_1 + \lambda (j-1)/2)\Gamma(
\lambda_2 + \lambda (j-1)/2)}.
$$
The generalized hypergeometric function of several variables \\
$_2F_1^{(\lambda)}(a,b;c;x_1,\dots,x_m)$  is the unique symmetric power series
solution of the partial differential equations
$$
x_j(1-x_j){\partial^2 F \over \partial x_j^2}
+ [ c -{2 \over \lambda}(m -1) - (a + b+1 -{2 \over \lambda} (m-1))x_j]
{\partial F \over \partial x_j}
$$
\begin{equation}
+{2 \over \lambda} \left [ \sum_{k=1 \atop k \not= j}^m {1 \over x_k - x_j}
\left ( x_j(1-x_j){\partial \over \partial x_j} - x_k (1-x_k){\partial \over
\partial x_k} \right ) \right ]F -abF = 0
\end{equation}
with the initial condition $_2F_1^{(\lambda)}=1$ when $x_1= \dots =x_m=0$
[9,10]
(when $m=1$ - single variable case - the differential equation satisfied by
the Gauss hypergeometric equation results).

We need to be able to compute the action of the operator $(T_{\{y\}})^j$ on
(2.12).
For this purpose we first write ${_2F_1}^{(q)}$ in integral form.
\subsection{An integral formula for ${_2F_1}^{(q)}$}

The following integral representation is due to Zan [10]:
$$
{_2F_1}^{(2/\lambda)}(a,\lambda_1+\lambda (n-1)/2;\lambda_1+\lambda_2 + \lambda
 (n-1);z_1,\dots,z_n) \hfill
 $$
$$
 = A_n (\lambda_1 ,\lambda_2 ,\lambda) \int_{I^n} {_1{\cal F}_0}^{(2 /
\lambda)}
 (a;z_1,\dots,z_n;s_1,\dots,s_n)
 $$
 \begin{equation}
 \times D_{\lambda_1,\lambda_2,\lambda}(s_1,\dots,s_n)ds_1 \dots ds_n
 \end{equation}
 where
 \begin{equation}
 D_{\lambda_1,\lambda_2,\lambda}(s_1,\dots,s_n) :=
 \prod_{l=1}^q s_l^{\lambda_1-1}(1 - s_l)^{\lambda_2-1}
 \prod_{1 \le j < k \le q} |s_k - s_j|^\lambda
 \end{equation}
 and
 $$
 { _1{\cal F}_0}^{(2/\lambda)}(a;z_1,\dots,z_n;s_1,\dots,s_n) \hfill
$$
\begin{equation}
:= \sum_{k=0}^\infty {1\over k!} \sum_{|\kappa|=k} [a]_\kappa^{(2/\lambda)}
{C_\kappa^{(2/\lambda)}(z_1,\dots,z_n) C_\kappa^{2/\lambda}(s_1,\dots,s_n)
\over C_\kappa^{(2/\lambda)}(1,\dots,1)}
\end{equation}
In (2.18) the second sum is over all partitions $(\kappa_1,\dots,\kappa_n)$
of $k$ into $n$ parts, $C_\kappa^{(2/\lambda)}$ denotes a suitably normalized
Jack polynomial [9], and
$$
[a]_\kappa^{(\alpha)} := \prod_{j=1}^n \left (a-{1\over \alpha}(j-1)\right )_
{\kappa_j}
\eqno (2.19a)
$$
where
$$
(a)_k := a(a+1) \dots (a+k-1).
\eqno (2.19b)
$$

\setcounter{equation}{19}

\noindent In general no formulas expressing ${_1{\cal F}_0}^{(2/\lambda)}$ in
terms of
simpler functions are known. However, an exception is the equal variable
case $z_1= \dots =z_n=z$ when [4]
\begin{equation}
_1{\cal F}_0^{(2/\lambda)}(a;z,\dots,z:s_1,\dots,s_n)
= \prod_{l=1}^n(1-zs_l)^{-a}
\end{equation}
The interval of integration for each variable $s_1,\dots,s_n$ given in [10]
is $I=[0,1]$.

The l.h.s. of (2.16) corresponds to the ${_2F_1}^{(q)}$ function in (2.12) if
we take
\begin{equation}
2/\lambda = q, \hspace{0.2cm} a=-N, \hspace{.2cm} n=q, \hspace{.2cm} \lambda_1
=1/q, \hspace{.2cm} \lambda_2=-N-1, \hspace{.2cm}\lambda_l = e^{2 \pi i y_l/L}
\end{equation}
in the former. However, we see from the definition (2.17) that in this case
the integrand in (2.16) is not integrable at $s_j$ due to the factor
$(1-v_j)^{-(N+2)}$, for each $j=1,\dots,q$.
In this circumstance, in the one variable case (Gauss hypergeometric function)
we know that the contour of integration $I$ can be shifted to
\begin{equation}
I=(-\infty,0]
\end{equation}
and the integration formula (2.16) is both well defined and satisfies (2.15).
We expect this property to carry over to the $m$-variable case, and indeed it
does
with the change of normalization
\begin{equation}
A_n(\lambda_1,\lambda_2,\lambda) \mapsto
(-1)^{\lambda_1n} A_n(\lambda_1,-\lambda (n-1)-\lambda_1 - \lambda_2 +
1,\lambda)
\end{equation}
as is shown in Appendix A.
	Thus we can use the integral representation (2.16) with parameters given by
(2.21),
interval of integration (2.22) and change of normalization (2.23).

\subsection{An equivalent operator}

Using the integral representation in (2.12) we see from (2.16) and (2.11) that
we must
calculate
\begin{equation}
\left( T_{\{y\}} \right)^j \left [ \prod_{l=1}^q
e^{-\pi i \rho y_l} {_1{\cal F}_0^{(q)}}(-N;e^{2\pi i y_1/L},\dots,e^{2\pi i
y_q/L};
s_1,\dots,s_q) \right] \Big |_{y_1= \dots y_q=x}
\end{equation}
To accomplish this task we recall [4] that $T_{\{y\}}$ has
$$
\prod_{l=1}^q e^{-\pi i \rho y_l} C_{\kappa}^{(q)}(e^{2\pi i y_1/L}
,\dots,e^{2\pi i
 y_q/L})
$$
as an eigenfunction, with eigenvalue $t_\kappa$ say, for each partition
$\kappa$. Hence, using (2.18), we see that (2.24) is equal to
\begin{equation}
e^{-\pi i\rho q x} \sum_{j=0}^\infty {1\over j!} \sum_{|\kappa| = j}
[a]_\kappa^{(2/\lambda)} (t_\kappa)^j
{C_\kappa^{(q)}(e^{2\pi i x /L},\dots,e^{2\pi i x /L})
C_\kappa^{(q)}(s_1,\dots,s_q) \over C_\kappa^{(q)}(1,\dots,1)}
\end{equation}

If we know introduce an operator $T'_{\{s\}}$ acting on the coordinates
$s_1,\dots,s_q$ such that
\begin{equation}
T'_{\{s\}}C_\kappa^{(q)}(s_1,\dots,s_q)=t_\kappa C_\kappa^{(q)}(s_1,\dots,s_q)
\end{equation}
we see from (2.18) that
\begin{equation}
\left( T_{\{s\}}' \right)^j\: e^{-\pi i\rho q x}{
{_1\cal F}_0^{(q)}}
(-N;e^{2\pi i x/L},\dots,e^{2\pi i x/L};s_1,\dots,s_q)
\end{equation}
is identical to (2.25), so we can replace (2.24) by the equivalent operation
(2.27). The construction of $T'_{\{s\}}$ is simple, due to the symmetry between
the
eigenfunctions $C_\kappa^{(q)}(z_1,\dots,z_q)$ and
$C_\kappa^{(q)}(s_1,\dots,s_q)$
exhibited in (2.18). In Appendix B we show that
$$
T'_{\{s\}} = -2 \left ({2 \pi \over L}
 \right )^2
\left [ {q \over 2} \sum_{j=1}^q s_j^2 {\partial^2 \over \partial s_j^2}
+ \sum_{j,k=1 \atop j \not= k}^q {s_j^2 {\partial \over \partial s_j}\over s_j
-s_k}
\right ] \hfill
$$
\begin{equation}
+\left (q(4 \pi^2 \rho /L \right ) - \left (2 \pi /L)^2 \right )
\sum_{j=1}^q s_j {\partial \over \partial s_j} + q (1-q)\left ({\pi \over L}
\right )^2N
\end{equation}
The calculation of (2.27) is straightforward since the special case of the
${_1{\cal F}}_0^{(q)}$ function therein can be evaluated according to (2.20).

Substituting the evaluation (2.20) in (2.27) and replacing (2.24) by the
resulting formula we thus have from (2.11), (2.12) and (2.16) (with the
further specifications made after (2.23)) the formula
$$
-i G_{N+1}((x,t),(0,0)) \hfill
$$
$$
= - C_{N,q} \, A_q(1/q,N+1/q,2/q) e^{-\pi i\rho q x} \sum_{j=0}^\infty
{(-it/\hbar)^j \over j!} \hfill
$$
$$
\times \int_{(-\infty,0]^q} ds_1 \dots ds_q \,
\left [\left( T_{\{s\}}' \right)^j \prod_{l=1}^q
(1 - e^{2 \pi i x/L} s_l)^N\right ] \hfill
$$
\begin{equation}
\times
 D_{1/q,-N-1,2/q}(s_1,\dots,s_q)
\end{equation}

\subsection{The thermodynamic limit}

{}From the explicit form (2.28), the action of the operator
 $\left( T_{\{s\}}' \right)^j$ in (2.29) can readily be computed using a
computer algebra package for given $N$ and $j$. However, many terms result
and it doesn't seem possible to sum the series in $j$. However, an
analytic calculation is possible in the large $N,L$ limit.
	From (2.28) we see that to leading order in $N$ with respect to its action on
the function

\begin{equation}
\prod_{l=1}^q (1 - e^{2 \pi i x/L} s_l)^N
\end{equation}
we can replace $\left( T_{\{s\}}' \right)^j$ by
\begin{equation}
\left (-q \left ({2 \pi \over L }\right )^2
\sum_{j=1}^q s_j^2 {\partial^2 \over \partial s_j^2}
+ \pi q \rho \left ({ 4 \pi \over L} \right)
\sum_{j=1}^q s_j{\partial \over \partial s_j}\right )^j
\end{equation}
The operator (2.31) gives terms which are $O(1)$ and $O(1/N)$ whereas the
remaining
part of $\left( T_{\{s\}}' \right)^j$ gives terms $O(1/N)$ only, which can
therefore
be ignored. The $O(1)$ action of (2.31) on (2.30) is readily computed, and we
conclude
$$
\left( T_{\{s\}}' \right)^j\prod_{l=1}^q (1 - e^{2 \pi i x/L} s_l)^N \hfill
$$
\begin{equation}
= \left ( -q(2 \pi \rho)^2 \sum_{j=1}^q {s_j^2 \over (1-s_j)^2} + {s_j \over
1-s_j} \right )^j\prod_{l=1}^q (1 - e^{2 \pi i x/L} s_l)^N + O(1/N)
\end{equation}
Substituting (2.32) in (2.29) we recognise that the sum over $j$ is simply
the power series for the exponential. Hence
$$
 -i G_{N+1}((x,t),(0,0)) \hspace{5cm}
$$
$$
\sim \:
- C_{N,q} \, A_q(1/q,N+1/q,2/q) e^{-\pi i\rho q x} \hfill
$$
$$
\times
\int_{(-\infty,0]^q} ds_1 \dots ds_q \,
\prod_{l=1}^q \exp \left [ {it/\hbar} \left (
{s_j^2 \over (1-s_j)^2}+{s_j \over 1-s_j}\right ) \right ]
(1 - e^{2 \pi i x/L} s_l)^N
$$
\begin{equation}
\times
 D_{1/q,-N-1,2/q}(s_1,\dots,s_q)
\end{equation}
We are now close to deriving the result (1.4).

To finish off the calculation, we first note from (2.13) and (2.14) that
\begin{equation}
\lim_{N\rightarrow \infty}C_{N,q} \, A_q(1/q,N+1/q,2/q)= B_q
\end{equation}
where $B_q$ is given by (1.4b). Next we change variables
\begin{equation}
{s_j \over 1-s_j} = -t_j
\end{equation}
in the integral (2.33) so that it reads
$$
-\int_{[0,1]^q} dt_1 \dots dt_q \,
\prod_{l=1}^q \exp \left [ {it/\hbar} (t_j^2-t_j) \right ]
(1 - t_j(1-e^{2 \pi i x/L}))^N
$$
\begin{equation}
\times
 D_{1/q,1/q,2/q}(t_1,\dots,t_q)
\end{equation}
But
\begin{equation}
\lim_{N,L\rightarrow \infty \atop N/L=\rho}
(1 - t_j(1-e^{2 \pi i x/L}))^N
= e^{2 \pi i t_j \rho x}.
\end{equation}
Substituting (2.37) in (2.36), changing variables
\begin{equation}
t_j=(v_j+1)/2
\end{equation}
and substituting the resulting expression in (2.33), we obtain the formula
(1.4) for the single-particle Green's function (1.3).

\vspace{.5cm}
\noindent Note added: Since completing this work,
the generalization of the formula (1.3) to all rational
values of the coupling $q$, together with a derivation using Jack polynomials,
has been announced by Ha [12].
\vspace{.5cm}

\noindent
ACKNOWLEDGEMENTS

\noindent
I thank Professor F.D.M. Haldane for sending me [1]. This work was supported by
the Australian Research Council.

\pagebreak
\noindent
{\bf Appendix A}
\setcounter{equation}{0}
\renewcommand{\theequation}{A\arabic{equation}}
\vspace{.5cm}

\noindent Here we derive the integral representation (2.16) with interval of
integration
(2.22) and change of normalization (2.23). We start with the transformation
formula [4,eq.(3.6)\footnote{this formula is erroneously missing $C$}]
$$
{_2F_1}^{(2/\lambda)}(a,b;c;t_1,\dots,t_m) \hspace{2cm}
$$
$$
= C \,{ _2F_1}^{(2/\lambda)}(a,b;a+b+1+\lambda(n-1)/2-c;1-t_1,\dots,1-t_m)
\eqno ({\rm A}1a)
$$
where
$$
C := C_m(a,b,c;\lambda):={A_n(b-\lambda(n-1)/2,c-b-\lambda(n-1)/2,\lambda)
\over A_n(b-\lambda(n-1)/2,c-b-a-\lambda(n-1)/2,\lambda)}
\eqno ({\rm A}1b)
$$
\setcounter{equation}{1}
which is valid for $a \in Z_{\le 0}$. Use of (A1) in (2.16) gives
$$
{_2F_1}^{(2/\lambda)}(a,\lambda_1+\lambda (n-1)/2;a+1-\lambda_2
 ;z_1,\dots,z_n) \hfill
$$
$$
 = {A_n (\lambda_1 ,\lambda_2 ,\lambda)
\over C_n(a,\lambda_1+\lambda(n-1)/2,\lambda_1+\lambda_2+\lambda(n-1);\lambda)}
\int_{[0,1]^n} {_1{\cal F}_0}^{(2 / \lambda)}
 (a;1-z_1,\dots,1-z_n;s_1,\dots,s_n)
 $$
 \begin{equation}
 \times D_{\lambda_1,\lambda_2,\lambda}(s_1,\dots,s_n)ds_1 \dots ds_n
 \end{equation}
where again it is assumed $a \in Z_{\le 0}$.

Next we change variables
\begin{equation}
s_i = - {u_i \over 1 - u_i} \hspace{1cm} (i=1,\dots,n).
\end{equation}
To do this we note
$$
D_{\lambda_1,\lambda_2,\lambda}(s_1,\dots,s_n)ds_1 \dots ds_n \hfill
$$
\begin{equation}
= (-1)^{\lambda_1n}
D_{\lambda_1,- [ \lambda_1+\lambda_2+\lambda(n-1)-1],\lambda}
(u_1,\dots,u_n)du_1 \dots du_n
\end{equation}
Furthermore we can make use of the transformation formula given by the
following
result

\vspace{.5cm}
\noindent
\underline{Lemma}

\noindent We have
$$
{_1{\cal F}_0}^{(2 / \lambda)} \hfill
 (a;1-z_1,\dots,1-z_n;-{u_1\over 1-u_1},\dots,-{u_n \over 1-u_n})
$$
\begin{equation}
=\prod_{j=1}^n(1-u_j)^a
{_1{\cal F}_0}^{(2 / \lambda)}
 (a;z_1,\dots,z_n;u_1,\dots,u_n)
\end{equation}

\vspace{.5cm}
\noindent \underline{Proof}

\noindent We know that [10]
\begin{equation}
{_2F_1}^{(2/\lambda)}(a,b;c;y_1,\dots,y_n)=
\prod_{j=1}^n(1-y_j)^{-a}
{_2F_1}^{(2/\lambda)}(a,c-b;c;-{y_1 \over 1-y_1},\dots,-{y_n \over 1-y_n})
\end{equation}
Expressing both sides of (A6) in terms of the integral representation (2.16)
and changing variables $s_j \mapsto 1-s_j$ on the r.h.s. shows
$$
\int_{[0,1]^n}\left [ {_1{\cal F}_0}^{(2 / \lambda)}
 (a;y_1,\dots,y_n;s_1,\dots,s_n) \right . \hfill
$$
$$
\left .
-\prod_{j=1}^n(1-y_j)^{-a}
{_1{\cal F}_0}^{(2 / \lambda)}
 (a;-{y_1 \over 1 -y_1},\dots,-{y_n \over 1 - y_n};1-s_1,\dots,1-s_n)\right ]
$$
\begin{equation}
 \times D_{\lambda_1,\lambda_2,\lambda}(s_1,\dots,s_n)ds_1 \dots ds_n = 0
\end{equation}
Since the terms in the square brackets are independent of $\lambda_1$ and
$\lambda_2$ while (A7) is identically zero for all ${\rm Re}(\lambda_1),
{\rm Re}(\lambda_2)>0$, we can conclude the combination of terms in the
square brackets vanishes identically. The identity (A6) then follows by
noting from (2.18) that in general
$$
{_1{\cal F}_0}^{(2 / \lambda)}
 (a;y_1,\dots,y_n;s_1,\dots,s_n)
$$
is unchanged by interchanging all the variables $y_1,\dots,y_n$ with the
variables
 $s_1,\dots,s_n$.

Substituting (A4) and (A5) in (A2), noting that in general
\begin{equation}
\prod_{j=1}^n(1-u_j)^a D_{\lambda_1,\lambda_2,\lambda}(u_1,\dots,u_n)
= D_{\lambda_1,\lambda_2+a,\lambda}(u_1,\dots,u_n),
\end{equation}
and replacing $\lambda_2$ by $a+1-(\lambda_1+\lambda_2+\lambda(n-1))$ shows
that
(A2) reduces to (2.16) with the substitutions (2.22) and (2.23), as required.

\vspace{.5cm}
\noindent
{\bf Appendix B}
\setcounter{equation}{0}
\renewcommand{\theequation}{B\arabic{equation}}

\noindent
Here we construct the operator $T'_{\{s\}}$ defined in subsection 2.3 by the
eigenvalue equation
\begin{equation}
T'_{\{s\}}C_\kappa^{(q)}(s_1,\dots,s_q)=t_\kappa C_\kappa^{(q)}(s_1,\dots,s_q),
\end{equation}
where the eigenvalue $t_\kappa$ is first to be calculated from the eigenvalue
equation
$$
T_{\{y\}}\prod_{l=1}^q e^{-\pi i \rho y_l} C_{\kappa}^{(q)}(e^{2\pi i y_1/L}
,\dots,e^{2\pi i y_q/L})
$$
\begin{equation}
=t_\kappa \prod_{l=1}^q e^{-\pi i \rho y_l} C_{\kappa}^{(q)}(e^{2\pi i y_1/L}
,\dots,e^{2\pi i y_q/L})
\end{equation}
with $T_{\{y\}}$ given by (2.5b).

To calculate $t_\kappa$ we recall that the Jack symmetric polynomial \\
$ C_{\kappa}^{(q)}(e^{2\pi i y_1/L}
,\dots,e^{2\pi i y_q/L}),$
which is labelled by a partition $\kappa = (\kappa_1,\dots,\kappa_m) $ of
non-negative integers such that
\begin{equation}
\kappa_1 \ge \kappa_2 \ge \dots \ge \kappa_m \hspace{1cm} {\rm and}
\hspace{1cm} \sum_{j=1}^m \kappa_j = k,
\end{equation}
is the unique (up to normalization) solution of the eigenvalue equation
[4;eq. (2.11) with $\alpha=m=q$]
$$
T_{\{y\}} C_{\kappa}^{(q)}(e^{2\pi i y_1/L}
,\dots,e^{2\pi i y_q/L})
$$
\begin{equation}
=\left ( -qe'_\kappa(q) + \left ({\pi \over L} \right )^2[qN+N(N-1)] \right )
C_{\kappa}^{(q)}(e^{2\pi i y_1/L}
,\dots,e^{2\pi i y_q/L})
\end{equation}
where
$$
e_\kappa'(q)={2 \over q}\left ({2\pi \over L} \right )^2(e_\kappa(q)+k/2)
\eqno ({\rm B}5a)
$$
$$
e_\kappa(q)=q\sum_{j=1}^q \kappa_j(\kappa_j-1)/2 - \sum_{q=1}^q(j-1)\kappa_j
+(q-1)k
\eqno ({\rm B}5b)
$$
\setcounter{equation}{5}
Using (B4) and the fact that $C_{\kappa}^{(q)}$ is homogeneous of order $k$,
which implies
$$
\sum_{j=1}^q {\partial \over \ \partial \theta_j}
C_{\kappa}^{(q)}(e^{2\pi i y_1/L}
,\dots,e^{2\pi i y_q/L})
$$
\begin{equation}
={2\pi i \over L}k
C_{\kappa}^{(q)}(e^{2\pi i y_1/L}
,\dots,e^{2\pi i y_q/L}),
\end{equation}
the l.h.s. of (B2) is readily computed, and we find
\begin{equation}
t_\kappa = 2 (2 \pi /L)^2 e_\kappa (q) + k[q(4 \pi^2 \rho /L) - (2 \pi /L)^2]
+q(q-1)(\pi /L)^2N
\end{equation}

To construct $T'_{\{s\}}$ with this eigenvalue acording to (B1), we recall [9]
that
\begin{equation}
\Delta_{\{s\}}C_\kappa^{(q)}(s_1,\dots,s_q)=e_\kappa(q)
C_\kappa^{(q)}(s_1,\dots,s_q),
\end{equation}
where
\begin{equation}
\Delta_{\{s\}}:={q \over 2} \sum_{j=1}^q s_j^2 {\partial^2 \over \partial
s_j^2}
+ \sum_{j,k=1 \atop j \not= k}^q {s_j^2 {\partial \over \partial s_j}\over s_j
-s_k}
\end{equation}
Furthermore, we can rewrite (B6) as
\begin{equation}
\sum_{j=1}^q s_j{\partial \over \partial s_j}
C_\kappa^{(q)}(s_1,\dots,s_q)=k C_\kappa^{(q)}(s_1,\dots,s_q)
\end{equation}
Hence
$$
\Big ( -2 \left ( 2 \pi /L\right )^2 \Delta_{\{s\}} +[ q(4 \pi^2 \rho /L) - (2
\pi /L)^2]
\sum_{j=1}^q s_j {\partial \over \partial s_j}
$$
$$
\left . +q(q-1)(\pi /L)^2 N \right )
C_\kappa^{(q)}(s_1,\dots,s_q)
$$
\begin{equation}
=t_\kappa C_\kappa^{(q)}(s_1,\dots,s_q)
\end{equation}
from which we read off that the explicit form of $T'_{\{s\}}$ is given by
(2.28).

\pagebreak

\noindent
{\bf References}

\vspace{.5cm}
\begin{description}
\item [1] F.D.M. Haldane, in {\it Proceedings of the 16th Taniguchi Symposium},
Kashikojima, Japan, 1993, eds. A. Okiji and N. Kawakami, Springer-Verlag, 1994.
\item [2] P.J. Forrester, Nucl. Phys. B {\bf 388} (1992) 671.
\item [3] P.J. Forrester, J. Stat. Phys. {\bf 72} (1993) 39.
\item [4] P.J. Forrester, Nucl. Phys. B {\bf 416} (1994) 377.
\item [5] P.J. Forrester, Phys. Lett. A {\bf 179} (1993) 127.
\item [6] F.D.M. Haldane and M.R. Zirnbauer, Phys. Rev. Lett. {\bf 71} (1993)
4055
\item [7] E.K.U. Gross, E. Runge and O. Heinonen, {\it Many Particle Theory},
(Adam Hilger, Bristol, 1991).
\item [8] B. Sutherland, Phys. Rev. A {\bf 4} (1971) 2019.
\item [9] J. Kaneko, SIAM J. of Math. Analysis{\bf 44} (1993) 1086.
\item [10] Y. Zan, Canad. J. of Math. {\bf 44} (1992) 1317.
\item [11] I.G. Macdonald, {\it Lecture Notes in Math.} vol. {1271}
(Springer, Berlin, 1987) 189.
\item [12] Z.N.C. Ha, "Exact Dynamical Correlation Functions of
Calogero-Sutherland
Model and One-dimensional Fractional Statistics", preprint, submitted Phys.
Rev. Lett.

\end{description}

\end{document}